\documentclass[%
 reprint,
superscriptaddress,
nobibnotes,
 amsmath,amssymb,
 aps,
prl,
showkeys
]{revtex4-2}

\usepackage{float}
\usepackage{graphicx}
\usepackage{dcolumn}
\usepackage{bm}
\usepackage{upgreek}
\usepackage{xcolor} 
\usepackage{soul,xcolor}
\usepackage{amsmath}
\usepackage{fancyhdr}
\usepackage{xr-hyper}

\usepackage[colorlinks=true,linkcolor=blue,urlcolor=blue,citecolor=blue]{hyperref}

\usepackage{hyperref}

\usepackage{pdfpages}

\makeatletter
\AtBeginDocument{\let\LS@rot\@undefined}
\makeatother

\setstcolor{orange}

\begin{document}

\preprint{APS/123-QED}

\title{Charge Exchange Dynamics in Cold Collisions of  $^{40}$CaH$^+$ and $^{39}$K}

\author{Swapnil Patel}
\altaffiliation{Current Address: Lawrence Berkeley National Lab, USA}
\affiliation{Duke Quantum Center, Duke University, Durham, NC 27701, USA}
\affiliation{Department of Physics, Duke University, Durham, NC 27708, USA}

\author{Dibyendu Sardar}
\affiliation{Faculty of Physics, University of Warsaw, Pasteura 5, Warsaw 02-093, Poland}

\author{Jyothi Saraladevi}
\altaffiliation{Current Address: IonQ Inc, USA}
\affiliation{Duke Quantum Center, Duke University, Durham, NC 27701, USA}
\affiliation{Department of Electrical and Computer Engineering, Duke University, Durham, NC 27708, USA}

\author{Michał Tomza}
\email{michal.tomza@fuw.edu.pl}
\affiliation{Faculty of Physics, University of Warsaw, Pasteura 5, Warsaw 02-093, Poland}

\author{Kenneth R. Brown}
 \email{kenneth.r.brown@duke.edu}
 \affiliation{Duke Quantum Center, Duke University, Durham, NC 27701, USA}
 \affiliation{Department of Physics, Duke University, Durham, NC 27708, USA}
\affiliation{Department of Electrical and Computer Engineering, Duke University, Durham, NC 27708, USA}
\affiliation{Department of Chemistry, Duke University, Durham, NC 27708, USA}

\date{\today}

\begin{abstract}

We report the observation of charge-exchange collisions between trapped calcium monohydride molecular ions ($^{40}$CaH$^+$) and ultracold potassium atoms ($^{39}$K) in a hybrid ion-atom trap. The measured charge-exchange rate coefficient is significantly suppressed relative to the Langevin rate constant for the system. We use \textit{ab initio} quantum-chemical calculations to model the (CaH-K)$^+$ complex in the ground and excited electronic states and to identify possible charge-exchange mechanisms. Our calculations rule out a direct non-radiative charge-exchange reaction and instead point to a radiative mechanism, but  do not quantitatively reproduce the measured rate, highlighting the need for a full-dimensional quantum dynamics treatment that includes vibrational motion and intermediate complex formation. Our work demonstrates that cold hybrid ion-atom platforms with molecular ions enable access to richer chemical complexity and collisional dynamics inaccessible in purely atomic systems.


\end{abstract}

\maketitle

The use of hybrid trap systems that combine ion and neutral traps, usually with the ability to precisely control the overlap of the two traps—has led to investigations of interactions between various combinations of atomic ions and neutral atoms. These studies have explored buffer gas cooling, molecular ion formation, impurity physics, and ultracold chemistry between the ions and the atoms~\cite{tomza2019cold, deiss2024cold}.

Simultaneously, techniques for trapping and controlling molecular ions—typically co-trapped and sympathetically cooled with atomic ions—have developed to allow precise preparation, manipulation, and coherent control of quantum states within the molecules~\cite{sinhal2023molecular, chou2017preparation, wolf2016non, sinhal2020quantum, lin2020quantum, chou2020frequency, liu2024quantum, holzapfel2025quantum}. The stability and long trapping times offered by ion traps, combined with the rich internal structure of molecular ions have facilitated their use in precision spectroscopy, fundamental physics, and quantum information processing~\cite{sinhal2023molecular, deiss2024cold}.

These two avenues have typically remained separate. There have been relatively few studies examining the combination of hybrid traps to investigate interactions between molecular ions and neutral atoms. Unlike atomic ions, molecular ions possess internal rotational and vibrational degrees of freedom. These modes can couple non-trivially to the translational motion during a collision, potentially leading to the processes that fundamentally alter reaction outcomes. Previous works probing these low temperature collisions include observing the evolution and the interplay of different collisional regimes by immersing a single molecular ion in a bath of ultracold Rb atoms~\cite{mohammadi2021life},  using inelastic collisions with ultracold Ca atoms to quench the vibrational degrees of freedom of BaCl\textsuperscript{+}~\cite{hudson2009method, rellergert2013evidence}, and investigations of ultracold chemical reactions. Hybrid systems have been employed in the study of associative detachment in molecular anions~\cite{deiglmayr2012reactive}, the synthesis of complex hypermetallic oxides~\cite{puri2017synthesis}, and investigations of charge transfer dynamics in homonuclear diatomics and polyatomic molecular ions~\cite{hall2012millikelvin, puri2019reaction, dorfler2019long, voute2023charge}. In many of these cases, the measured phenomena often deviated from classical theory predictions, and thus state-of-the-art quantum chemistry calculations were employed to explain the underlying processes and refine theoretical quantum chemistry models~\cite{KarmanNP24}. 

In this Letter, we use a hybrid setup to study the interaction between heteronuclear diatomic molecular ions and neutral atoms with similar mass. We observe charge exchange (CE) between sympathetically trapped \textsuperscript{40}CaH\textsuperscript{+} molecular ions and ultracold neutral \textsuperscript{39}K atoms and characterize this interaction experimentally and theoretically. The measured reaction rate coefficient analyzed alongside quantum-chemical modeling points towards a radiative mechanism and highlights the importance of molecular vibration and intermediate complex formation in molecular ion-atom charge-exchange reactions.

\begin{figure*}[t]
    \begin{tabular}{cc}
        \raisebox{1.9 cm}{\textbf{a}}
        \hspace{0.50cm}\includegraphics[width=1.25\columnwidth]{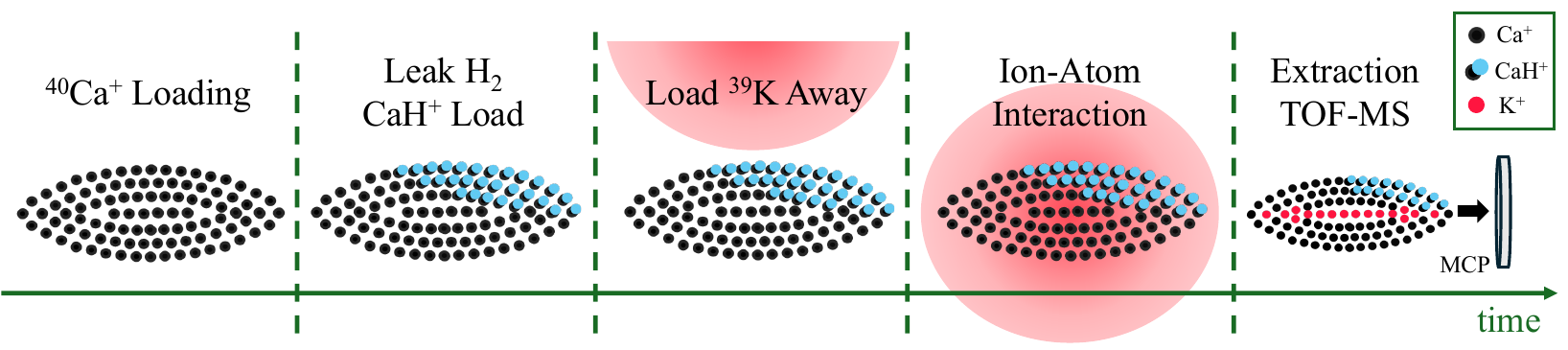}\hspace{0.50cm}
    \end{tabular}
    \begin{tabular}{cc}
        \raisebox{5 cm}{\textbf{b}}
        \hspace{-0.125cm}\includegraphics[width=0.29\linewidth]{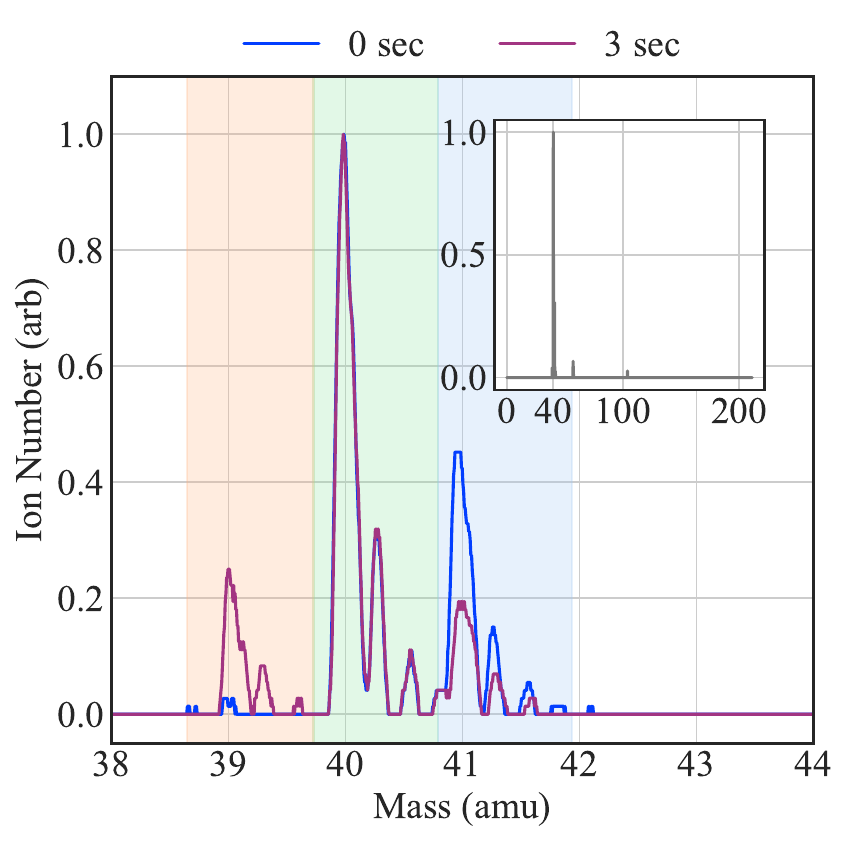} 
        
        \raisebox{5 cm}{\textbf{c}}
        \hspace{0.0cm}\includegraphics[width=0.35\linewidth]{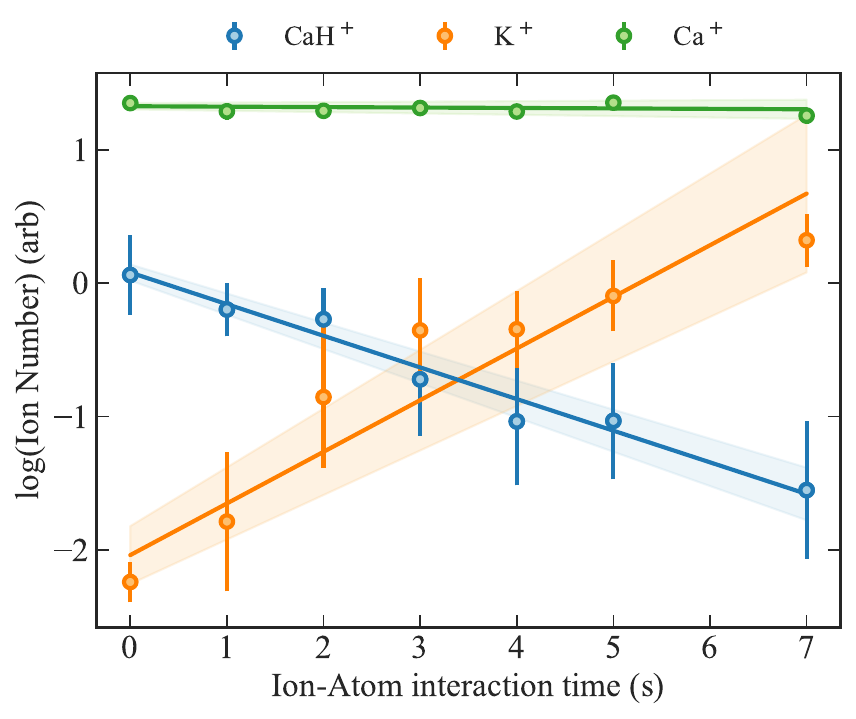}
    \end{tabular}
    \caption{(a) Experimental sequence for interactions between CaH\textsuperscript{+} and K. The ion-atom interaction time and the \textsuperscript{39}K MOT parameters are varied to characterize the CE reaction. (b) Example TOF-MS signal indicating the presence of mass at 39, 40, 41 amu corresponding to K\textsuperscript{+}, Ca\textsuperscript{+}, CaH\textsuperscript{+} ions in the trap. The blue and the red curve show the signal after ion-atom hold time of 0 and 3 seconds respectively. The inset shows the complete signal range. (c) Integrated MS signal as a function of ion-atom hold time for K\textsuperscript{+}, Ca\textsuperscript{+}, and CaH\textsuperscript{+} with pseudo-first order rate fits (solid lines). Error bars correspond to one standard deviation. The MOT density and average P-state population are $4.97(42) \times 10^8$ cm\textsuperscript{-3}  and $13.6(2)\%$ respectively.}
    \label{fig:expsetup}
\end{figure*}

The experimental setup, depicted in Fig.~\ref{fig:expsetup}(a) and detailed in Ref.~\cite{jyothi2019}, is a hybrid ion-atom trap integrated with a time-of-flight mass spectrometer (TOF-MS). In a typical experimental run for this study, we trap and Doppler cool about 120 calcium ions (\textsuperscript{40}Ca\textsuperscript{+}) in a linear four-rod Paul trap, with an RF drive frequency of $\Omega_{rf} = 2\pi \times 1.7 \, \mathrm{MHz}$, and a trap axis to electrode distance of $r_0 = 9.6 \, \mathrm{mm}$. Calcium monohydride (\textsuperscript{40}CaH\textsuperscript{+}) ions are produced by leaking H\textsubscript{2} gas into the chamber through a piezo-electric leak valve. The gas is leaked for 4 minutes at pressures of about $ 4 \times 10^{-8}$ torr (base pressure: $9 \times 10^{-10}$ torr) and we typically convert about 25\% of the calcium ions to calcium hydride ions. The molecular ions are co-trapped and sympathetically cooled by the laser-cooled calcium ions~\cite{rugango2015}. We assume that the vibrational and rotational degrees of freedom thermalize with the room temperature black-body radiation, and that sympathetic cooling by Ca\textsuperscript{+} ions does not quench these internal states~\cite{bertelsen2006}. In our case, \textsuperscript{40}CaH\textsuperscript{+} is in its electronic and vibrational ground state at room temperature, but the rotational population is distributed over the first 15 rotational states (peak population in $J=4$ with approximately 13\% population)~\cite{patel2025precise}. 

A 3D magneto-optical trap (MOT) designed to trap potassium atoms (\textsuperscript{39}K) completes our hybrid setup. The MOT is generated by 767$\,$nm trapping beams, which are retro-reflected from three orthogonal input beams. A pair of copper coils in an anti-Helmholtz configuration provides the necessary magnetic field gradient, with the field's center designed to coincide with the center of the ion trap. An additional pair of coils, in Helmholtz configuration, allows for shifting the center of the magnetic field gradient in the vertical direction.

After loading a Coulomb crystal consisting of Ca\textsuperscript{+}-CaH\textsuperscript{+} ions, we load the \textsuperscript{39}K MOT away from the center of the ion trap. The atoms are then overlapped with the ion crystal for a variable amount of time by switching off the additional Helmholtz field. Finally, the MOT is switched off and the resulting cations in the trap are ejected to the time-of-flight mass spectrometer. The resolution of the TOF-MS is better than 1 amu and the integrated signal linearly corresponds to the number of ions in the trap. An example mass spectrometer signal zoomed in to the relevant mass region is shown in Fig.~\ref{fig:expsetup}(b).

During the overlap, the collision energy is dictated by the steady-state kinematics of the trapped ions, as their driven micromotion energy is much higher than that of the atoms. Due to the higher mass of CaH\textsuperscript{+} relative to Ca\textsuperscript{+}, the molecular ions occupy the outer boundary of the 3D crystal. Based on our trap parameters, we estimate an effective collision energy of $E_{coll}/k_{B}$ $\approx 1$ K ($\approx 0.7 \, \mathrm{cm}^{-1})$. This is in agreement with our measurement of the effective temperature, obtained by fitting a Voigt profile to the fluorescence spectroscopy of the Ca\textsuperscript{+} 397 nm transition in the 3D Coulomb crystal.

\begin{figure}[t]
    \centering
    \includegraphics[width=0.9\linewidth]{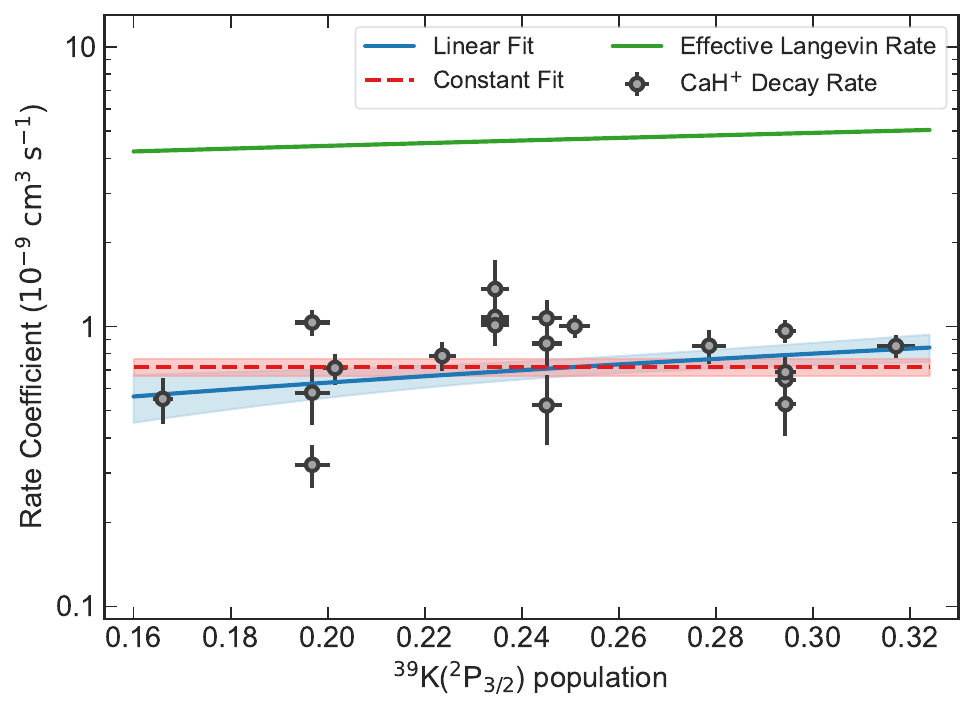}
    \caption{Measured charge exchange rate coefficients as a function of the average excited state population of \textsuperscript{39}K atoms (\textsuperscript{2}P\textsubscript{3/2}). Two model fits are shown: a constant fit (dashed red line) assuming no internal state dependence, and a linear fit (solid blue line) allowing for state dependence. The fit is represented by $k_{eff} = (k_P - k_S)p + k_S$, where $p$ is the average excited state population of the K atoms, and $k_S$ and $k_P$ are the rates for the ground and the excited state. The shaded regions represent $1\sigma$ confidence bands.
    The effective Langevin rate (solid green line) represents a weighted average of the S- and P-state Langevin rates.}
    \label{fig:cah-k_data_bothfits}
\end{figure}

Before exploring the interaction between \textsuperscript{40}CaH\textsuperscript{+} and \textsuperscript{39}K, we must first address the formation of \textsuperscript{39}K\textsuperscript{+} from the charge exchange between \textsuperscript{40}Ca\textsuperscript{+} and \textsuperscript{39}K. We have previously experimentally and theoretically characterized the above reaction, noting that the charge exchange predominantly occurs from the excited state of the calcium ions~\cite{li2020}. For experiments with the molecular ions and the atoms, we minimize this background charge exchange process by minimizing the excited state population of the calcium ions.

The sequence described above, and illustrated in Fig.~\ref{fig:expsetup}(a), is repeated for varying ion-atom hold times while maintaining the \textsuperscript{39}K MOT parameters constant. We observe charge exchange through the decay of \textsuperscript{40}CaH\textsuperscript{+} ions and the corresponding growth of \textsuperscript{39}K\textsuperscript{+} ions (Fig.~\ref{fig:expsetup}(c)). The ion numbers as a function of ion-atom hold times, for each of the three species, is fit to a pseudo-first-order rate equation, $\ln [N_{x}(t) / N_{x}(0)] = -k_{x} t$ where $N_{x}(t)$ is the number of ions of species $x$ at time $t$ and $k_{x}$ is the corresponding fit rate~\cite{hall2012millikelvin}.

To distinguish the contributions of the ground (\textsuperscript{2}S\textsubscript{1/2}) and the excited (\textsuperscript{2}P\textsubscript{3/2}) states of \textsuperscript{39}K to the observed charge exchange process, we repeat the measurements for various combinations of \textsuperscript{39}K atom number density and average excited state population in the MOT. This is achieved by varying the trapping laser intensity and the anti-Helmholtz field strength used to trap the atoms. For each of the given combinations, the total rate coefficients were obtained as $k_\text{eff} = k_\text{CaH/K} / n_\text{K}$, where $k_\text{CaH/K}$ are the pseudo-first-order rates and $n_\text{K}$ is the corresponding atom number density of the \textsuperscript{39}K MOT. 

\begin{figure}[t!]
\centering
\includegraphics[width=0.9\linewidth]{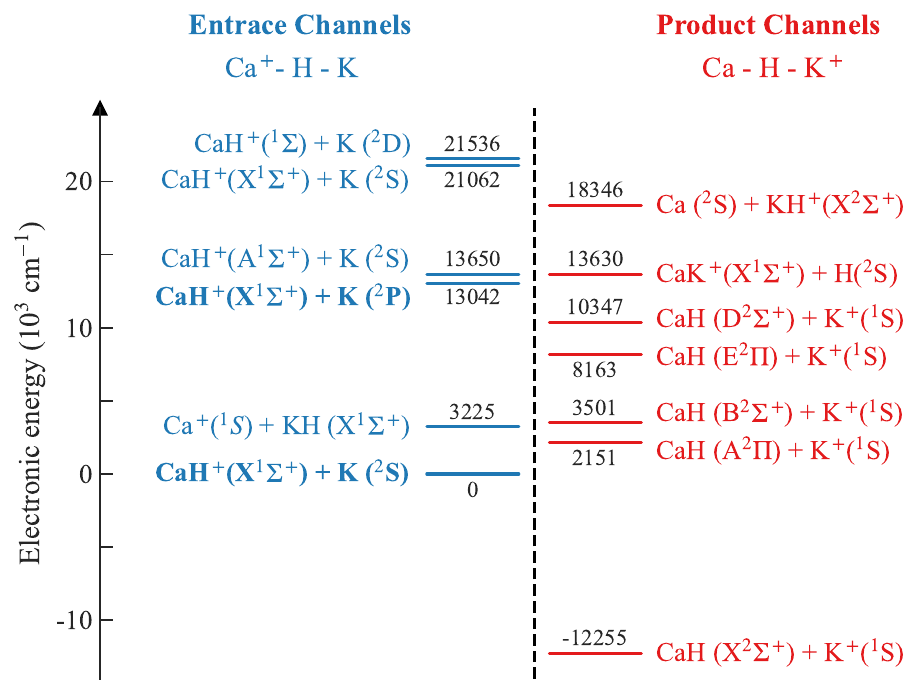}
\caption{Energetics of charge exchange collisions. Asymptotic energies of possible atom-molecule entrance and product channels below 22,000$\,$cm$^{-1}$ for the (Ca-H-K)$^+$ system. Channels on the left correspond to the arrangements with a charge formally assigned to Ca and on the right -- to K. The channels in bold on the left represent the energetically accessed initial states.}
\label{fig:channels}
\end{figure}

Fig.~\ref{fig:cah-k_data_bothfits} shows the dependence of the charge-exchange rate coefficient on the average excited state population ($p$) of the potassium atoms in the MOT. We fit the data using two models. A constant fit (dashed red line in Fig.~\ref{fig:cah-k_data_bothfits}) assuming no internal state dependence yields a rate coefficient of $k = 0.72 (5) \times 10^{-9} $ cm\textsuperscript{3} s\textsuperscript{-1}, and a linear fit (solid blue line in Fig.~\ref{fig:cah-k_data_bothfits}) based on a state-dependent model $k_{eff} = k_S(1 - p) + k_Pp$, yields a ground state rate $k_S = 0.29 (27) \times 10^{-9} $ cm\textsuperscript{3} s\textsuperscript{-1}, and an excited state rate $k_P = 1.99 (81) \times 10^{-9} $ cm\textsuperscript{3} s\textsuperscript{-1}.

We note that both models result in large reduced chi-squared values (5.61 for the constant fit and 5.20 for the linear fit). This indicates that both fits describe the data comparably, with large scatter compared to the statistical uncertainties. The modest statistical preference towards the linear fit provides, at best, weak evidence for an internal-state dependence.

The ground state rate is an order of magnitude smaller than the classical Langevin rate coefficient ($k_\text{L} (S_{1/2})=3.44 \times 10^{-9} \mathrm{\, cm^3 \, s^{-1}}$), while the excited state rate approaches the Langevin limit within a factor of 2.5 for 100\% P-state population ($k_\text{L}(P_{3/2})=4.99 \times 10^{-9} \mathrm{\, cm^3 \, s^{-1}}$)~\cite{langevin1905fundamental}.  Even with the Langevin rate scaled by the excited state-population, as shown by the green curve in Fig.~\ref{fig:cah-k_data_bothfits}, the experimental values remain significantly suppressed.

The only energetically possible outcome of a cold reactive collision between ground-state CaH$^+$ and K is a charge exchange resulting in ground-state CaH and K$^+$. When a potassium atom is electronically excited, other product channels become energetically accessible, including several channels with CaH in excited electronic states and a proper chemically reactive channel leading to Ca$^+$ and KH. All relevant channels are collected in Fig.~\ref{fig:channels}. Charge exchange processes can proceed via four main pathways: direct chemical reaction on a single potential energy surface (PES), non-radiative transfer driven by non-adiabatic coupling, stimulated radiative transitions via photon absorption, or spontaneous radiative emission.

\begin{figure}[tb!]
\centering
\includegraphics[width=0.9\linewidth]{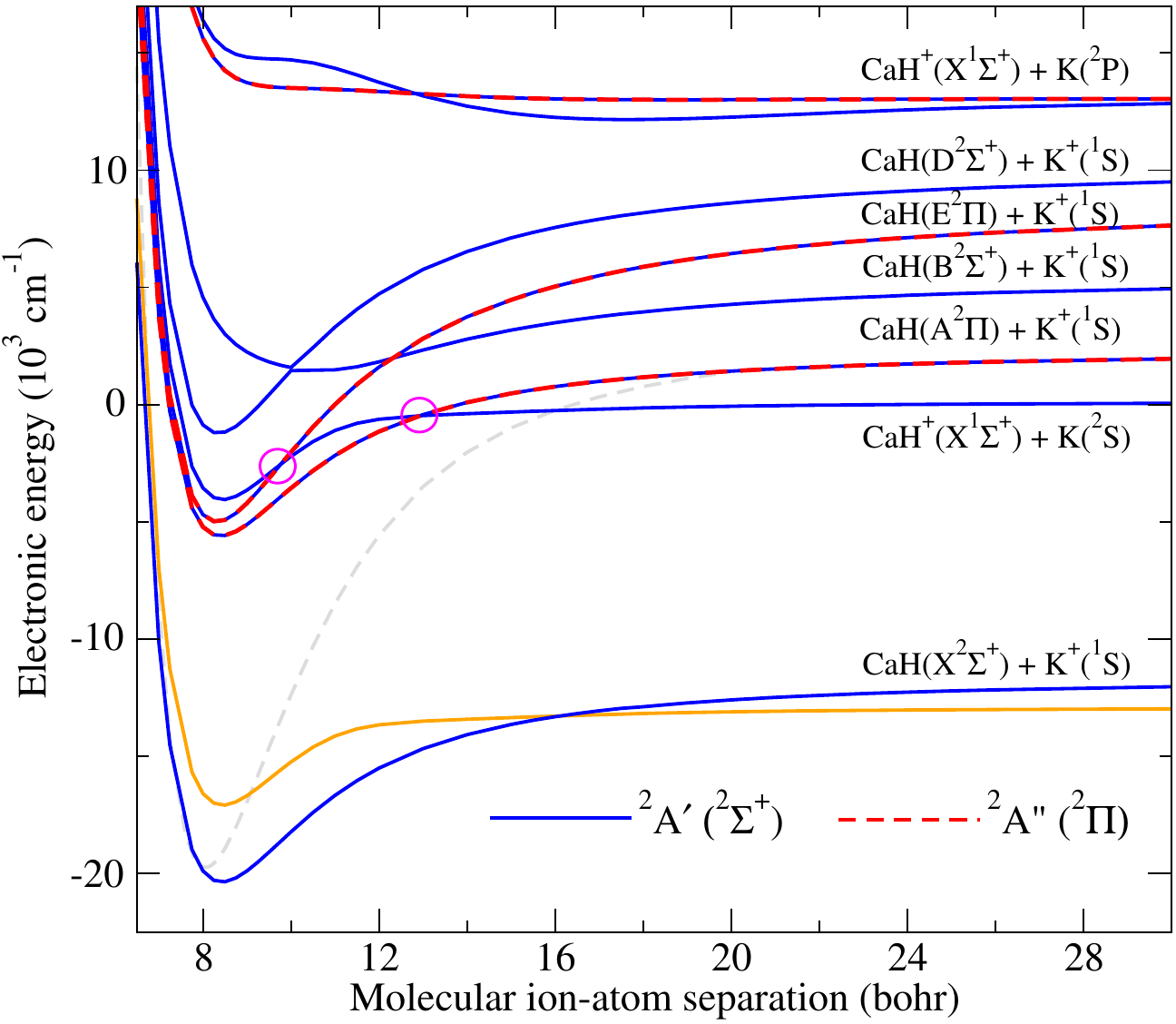}
\caption{One-dimensional cut through the potential energy surfaces for the ground
and excited electronic states of the (CaH-K)$^+$ system in the linear geometry. Pink circles highlight the crossings between the entrance channel with higher-energy states of different symmetry. The solid orange curve represents the entrance channel potential shifted by the photon energy ($h\nu \approx 13,000\,$cm$^{-1}$), illustrating the resonance with the exit channel. The dashed gray line represents the ground-state interaction for Ca$^+$ with KH (note a different definition of the $x$-axis for this curve resulting in an artificial crossing with the lowest PES for CaH+K$^+$).}
\label{fig:PESs}
\end{figure}

We carefully calculated and explored representative 1D and 2D cuts of the potential energy surfaces for the lowest electronic states of the (CaH-K)$^+$ system to evaluate if direct non-radiative charge transfer on a single PES or through PES crossings is possible. In such a situation, charge exchange rates close to the Langevin limit could be expected, as in N$_2^+$+Rb and O$_2^+$+Rb~\cite{dorfler2019long} or N$_2$H$^+$+Rb\cite{voute2023charge}. The exemplary 1D cut through the PESs of our system in the linear (CaHK)$^+$ geometry is presented in Fig.~\ref{fig:PESs}. This geometry was selected because it exhibits the strongest interaction and the deepest potential wells in the entrance and higher-lying channels, and therefore most clearly illustrates the relevant PES topology and crossings between potential energy surfaces. Additional cuts for other representative geometries are provided in the Supporting Information. The entrance CaH$^+$(X$^1\Sigma^+$)+K($^2$S) channel is energetically well separated from the exit CaH(X$^2\Sigma^+$)+K$^+$($^1$S) configuration. In our extensive scans of the corresponding PESs for different geometries and symmetries, we have not found any direct crossing between the entrance and exit PESs. The entrance channel crosses with higher-energy states, but we also have not found the path from the entrance channel to the exit channel through crossings between surfaces of different symmetries ($^1\Sigma$ and $^2\Pi$ for the linear configurations or $^2A'$ and $^2A''$ for other geometries, see Fig.~\ref{fig:PESs}). 
Finally, we have probed atom-exchange geometries (CaH$^+$+K$\to$Ca$^+$+KH), but no path to the exit channel has been found. We conclude that direct chemical reaction for colliding CaH$^+$ and K in their ground electronic state is improbable, limiting the importance of stereodynamics relevant for other systems~\cite{dorfler2019long,voute2023charge}. In our calculations, we also have not observed any indication of avoided crossings between the entrance and exit channels that would facilitate non-radiative transfer.

The cooling lasers present in the experiment could, in principle, be responsible for stimulated radiative charge exchange via photon absorption. Although the light used for laser cooling of K atoms is energetically close, it is too energetic to directly stimulate charge exchange (see the orange line in Fig.~\ref{fig:PESs}). Indirect processes, such as photoassociation to ground-state triatomic CaHK$^+$ followed by excitation to short-lived states correlating to the CaH(A$^2\Pi^+$)+K$^+$($^1$S) asymptote and spontaneous decay to CaH(X$^2\Sigma^+$)+K$^+$($^1$S), are in principle possible~\cite{mohammadi2021life}. During our measurements, the 767 nm laser intensity was varied across more than three-fold range (from approximately 90 to 310 $\mathrm{W/m}^2$). As detailed in the Supporting Information, plotting the measured rate coefficient directly against laser intensity reveals no significant scaling with the applied laser intensity. This lack of explicit intensity dependence indicates that the direct laser-stimulated processes make a negligible contribution to the measured rate.

Spontaneous radiative charge exchange and association between ground-state atomic ions and atoms is known to be relatively slow, with measured rate coefficients typically in the range of $10^{-4}-10^{-3}$ $k_\text{L}$ (the Langevin rate coefficient)~\cite{HallPRL11,RatschbacherNatPhys12,JogerPRA17,SaitoPRA17,SikorskyNatCom18}. The rate coefficient for Ca$^+$+K$\to$Ca+K$^+$ collisions was also calculated to be below $10^{-4}\,k_\text{L}$~\cite{ZrafiNJP20}. Such a radiative process has not been previously observed for any molecular ion-atom system, but was theoretically predicted for the N$_2^+$+Rb combination to be equally slow~\cite{GianturcoPCCP19}. Our present calculation of spontaneous radiative charge exchange and association within the rigid-rotor and infinite-order-sudden approximations on a single entrance $^2A'$ PES yields a rate coefficient of order $10^{-3}\,k_\text{L}$ that is more than two orders of magnitude below. Given the approximate nature of the present calculations, we report only an order-of-magnitude theoretical estimate. While the calculated rate may increase due to enhanced wavefunction amplitude at crossings of the entrance PES with neglected higher-lying electronic states in a multichannel treatment (see Fig.~\ref{fig:PESs}), such effects are unlikely to account for the observed mismatch.

The discrepancy between the measured reaction rate and these standard mechanisms implies that additional dynamics---beyond those captured by our current model---may play a role in enhancing the reaction rate. One possibility is that the reaction is mediated by the formation of long-lived intermediate complexes, which can decay radiatively or through collisions with another atom. In neutral molecule collisions, such complexes have been observed to possess extremely long lifetimes, thereby significantly modifying the collisional dynamics~\cite{NicholsPRX22,BauseJPCA23,JachymskiPRA22}. A similar mismatch between experimental results and rigid-rotor scattering predictions was recently analyzed in ultracold Rb+KRb collisions~\cite{LiuNChem25}. Our present scattering calculations rely on the rigid-rotor, single-surface, and infinite-order sudden approximations. The current model is thus unable to fully capture the full wavepacket dynamics associated with such complexes and highlights an inherent distinction between atomic and molecular systems modeling that must be included in future theoretical studies.

We have observed charge exchange collisions between trapped \textsuperscript{40}CaH\textsuperscript{+} and ultracold \textsuperscript{39}K in a hybrid ion-atom trap. The reaction rate is approximately a factor of five lower than the Langevin collision rate. This measurement was motivated by and stands as a prerequisite for our objective of sympathetically cooling the rotational motion of the molecular ion through inelastic quenching collisions with the ultracold atoms~\cite{hudson2009method}. Although the charge exchange channel represents a loss mechanism for the molecular ions, the moderate rate indicates that sympathetic cooling remains viable. Experimentally, we aim to utilize the black-body thermometry based on rotational spectroscopy of \textsuperscript{40}CaH\textsuperscript{+} to quantify the rate of quenching collisions and to prepare \textsuperscript{40}CaH\textsuperscript{+} in its rovibrational ground state~\cite{patel2025precise}. These capabilities will enable controlled studies of internal state dynamics and thermalization in molecular ion–neutral atom systems.

The measured charge-exchange rate coefficients also challenge modern quantum-chemical theory to reproduce observed values. The modeling presented here indicates the importance of molecular vibration and intermediate complex formation. Future theoretical work should move beyond the rigid-rotor approximation, carefully evaluate the enhancement of intermediate complex lifetimes in a trap, and different spin couplings that may link the present results to the so-called sticky collisions in ultracold neutral molecular gases~\cite{BauseJPCA23,NicholsPRX22}. Experimentally, future studies utilizing 'dark' traps - such as a dark-spot MOT, chopped optical dipole traps-could definitively decouple laser stimulated pathways from intrinsic collisional dynamics. 

\section{\label{sec:theorymethods}Theoretical Methods}
We employ \textit{ab initio} quantum-chemical methods to model and understand charge-exchange paths and rate constants. First, potential energy surfaces (PESs) for ground and excited electronic states are calculated and inspected using advanced electronic structure methods. Next, radiative charge-exchange rates are approximated with the infinite‐order sudden approximation.    

We apply the internally contracted multireference configuration interaction method restricted to single and double excitations (MRCISD)~\cite{WernerJCP88} to calculate two-dimensional potential energy surfaces for the ground and excited electronic states. Although MRCISD is neither size-extensive nor size-consistent~\cite{Szalay2012}, we use uncorrected MRCISD energies to obtain smooth and qualitatively reliable potential-energy-surface topologies, avoiding possible non-smooth state-dependent shifts from Davidson corrections near avoided crossings and conical intersections~\cite{Matsika2021, Park2020}. A large active space constructed from the highest occupied and eight lowest unoccupied molecular orbitals of CaH, along with the 4s, 4p, 5s, 3d, and 5p atomic orbitals of potassium, is used. Additionally, the coupled cluster method restricted to single, double, and noniterative triple excitations [CCSD(T)]~\cite{KnowlesJCP93} is applied to study the lowest electronic states of different symmetries within the $C_s$ and $C_{2v}$ point groups. 

Potential energy surfaces are calculated for several configurations, including one-dimensional (1D) cuts with CaH described as a rigid rotor and different atom-molecule orientations, two-dimensional (2D) cuts within the rigid-rotor approximation or for the fixed CaHK angle, and three-dimensional (3D) optimizations of the equilibrium geometries.    

Electronic orbitals are constructed using the augmented correlation-consistent polarized weighted core-valence quintuple-$\zeta$ quality basis set (aug-cc-pwCV5Z-PP~\cite{hill2017gaussian} for Ca and K and aug-cc-pV5Z~\cite{DunningJCP89} for H). We include the scalar relativistic effect in Ca and K using the small-core relativistic energy-consistent pseudopotential (ECP10MDF~\cite{lim2005all}) to replace the inner-shell electrons. Basis set superposition errors are corrected using the counterpoise correction. Electronic structure calculations are performed with the \textsc{Molpro}~\cite{WernerJCP20,Molpro,MOLPRO-WIREs} package of \textit{ab initio} programs.

The spontaneous charge-exchange and radiative association rate coefficients are computed using the Fermi's golden rule-type expressions based on the Einstein coefficients between continuum and continuum/bound nuclear wave functions of relevant electronic states as implemented in Refs.~\cite{TomzaPRA15a,ZrafiNJP20}. The infinite-order sudden (IOS) approximation is employed, which reduces the problem to one-dimensional scattering calculations for different \textsuperscript{40}CaH$^+$ orientations described within the rigid-rotor approximation~\cite{TsienJCP73,dorfler2019long}.
\\

\textit{Acknowledgments---}
This work is supported by the Army Research Office (W911NF-21-1-0346), the European Union (ERC, 101042989--QuantMol), and the Poland’s high-performance computing infrastructure PLGrid (HPC Center: ACK Cyfronet AGH, PLG/2024/017844). 

\bibliography{adco}

\clearpage

\onecolumngrid 

\includepdf[pages=1]{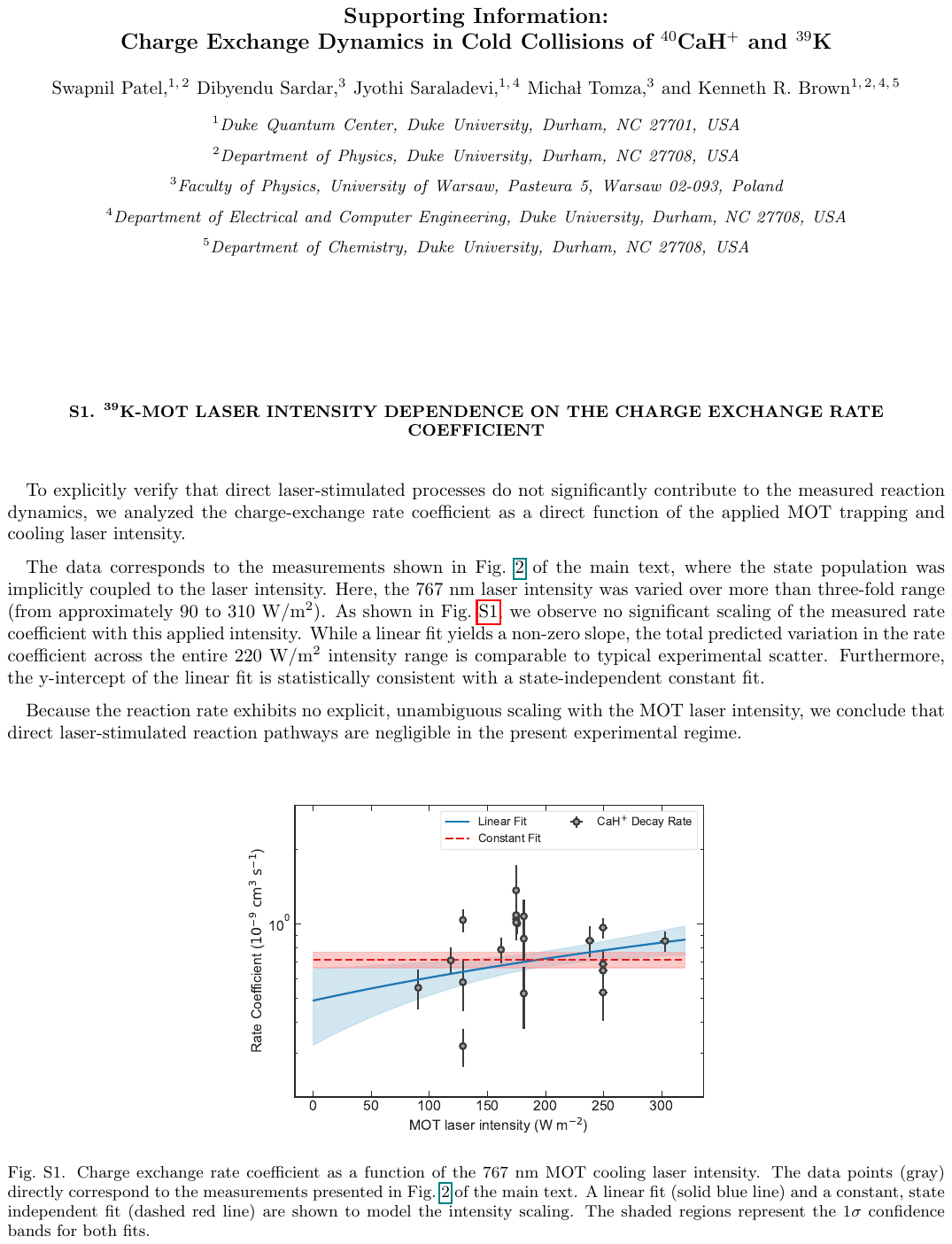}
\includepdf[pages=2]{supplement.pdf}

\end{document}


\title{Supporting Information: \\
Charge Exchange Dynamics in Cold Collisions of  $^{40}$CaH$^+$ and $^{39}$K}

\author{Swapnil Patel}
\affiliation{Duke Quantum Center, Duke University, Durham, NC 27701, USA}
\affiliation{Department of Physics, Duke University, Durham, NC 27708, USA}

\author{Dibyendu Sardar}
\affiliation{Faculty of Physics, University of Warsaw, Pasteura 5, Warsaw 02-093, Poland}

\author{Jyothi Saraladevi}
\affiliation{Duke Quantum Center, Duke University, Durham, NC 27701, USA}
\affiliation{Department of Electrical and Computer Engineering, Duke University, Durham, NC 27708, USA}

\author{Michał Tomza}
\affiliation{Faculty of Physics, University of Warsaw, Pasteura 5, Warsaw 02-093, Poland}

\author{Kenneth R. Brown}
 \affiliation{Duke Quantum Center, Duke University, Durham, NC 27701, USA}
 \affiliation{Department of Physics, Duke University, Durham, NC 27708, USA}
\affiliation{Department of Electrical and Computer Engineering, Duke University, Durham, NC 27708, USA}
\affiliation{Department of Chemistry, Duke University, Durham, NC 27708, USA}


\maketitle

\onecolumngrid

\section{S1. \textsuperscript{39}K-MOT Laser Intensity Dependence on the Charge Exchange Rate Coefficient}

To explicitly verify that direct laser-stimulated processes do not significantly contribute to the measured reaction dynamics, we analyzed the charge-exchange rate coefficient as a direct function of the applied MOT trapping and cooling laser intensity. 

\begin{figure}[]
    \centering
     \includegraphics[width=0.5\linewidth]{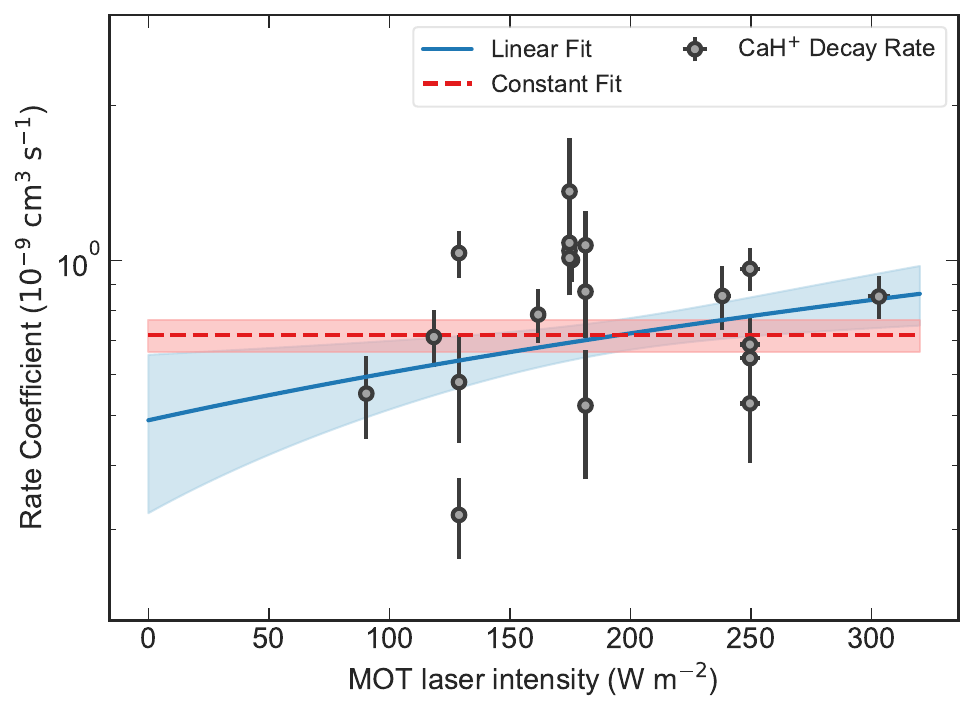}
    \caption{Charge exchange rate coefficient as a function of the 767 nm MOT cooling laser intensity. The data points (gray) directly correspond to the measurements presented in Fig.~\ref{fig:cah-k_data_bothfits} of the main text. A linear fit (solid blue line) and a constant, state independent fit (dashed red line) are shown to model the intensity scaling. The shaded regions represent the $1\sigma$ confidence bands for both fits.}
    \label{fig:laserintensity}
\end{figure}

The data corresponds to the measurements shown in Fig.~\ref{fig:cah-k_data_bothfits} of the main text, where the state population was implicitly coupled to the laser intensity. Here, the 767 nm laser intensity was varied over more than three-fold range (from approximately 90 to 310 $\mathrm{W/m^2}$). As shown in Fig.~\ref{fig:laserintensity}, we observe no significant scaling of the measured rate coefficient with this applied intensity. While a linear fit yields a non-zero slope, the total predicted variation in the rate coefficient across the entire 220 $\mathrm{W/m^2}$ intensity range is comparable to typical experimental scatter. Furthermore, the y-intercept of the linear fit is statistically consistent with a state-independent constant fit. 

Because the reaction rate exhibits no explicit, unambiguous scaling with the MOT laser intensity, we conclude that direct laser-stimulated reaction pathways are negligible in the present experimental regime. 

\pagebreak

\section{S2. One-dimensional cuts of potential energy surfaces for (CaH-K)$^+$}

We present one-dimensional cuts through the potential energy surfaces for the ground
and excited electronic states of the (CaH-K)$^+$ system for additional atom-molecule orientations, calculated as described and presented in Fig.~\ref{fig:PESs} of the main text. For nonlinear orientations, convergence issues were encountered for some high-lying electronic states, therefore the highest $^2A'$ state is missing in panel (b).

Three representative geometries are shown: $\theta = 0^\circ$, corresponding to the
linear (HCaK)$^+$ configuration; $\theta = 90^\circ$, corresponding to the T-shaped
configuration; and $\theta = 170^\circ$, corresponding to a slightly bent (CaHK)$^+$
configuration close to the linear geometry presented in Fig.~\ref{fig:PESs} of the main text. The
$\theta = 170^\circ$ cut illustrates how a small deviation from linearity and the
associated lowering of symmetry affect the PES topology, whereas the $\theta = 0^\circ$
and $\theta = 90^\circ$ cuts represent distinct approach geometries, for which some PESs
become predominantly repulsive and fewer crossings are observed. The entrance channel,
CaH$^+$(X$^1\Sigma^+$)+K($^2$S), remains well separated from the charge-exchange exit
channel, CaH(X$^2\Sigma^+$)+K$^+$($^1$S), for all three geometrical configurations.
No direct curve crossings are observed between the entrance- and exit-channel PESs in
any of these configurations. Moreover, for both linear and bent geometries, no direct
crossings are found within the classically allowed region between the entrance-channel
PES and higher-lying states that could provide a reactive pathway to the exit channel.

\begin{figure}[b]
\includegraphics[width=0.36\linewidth]{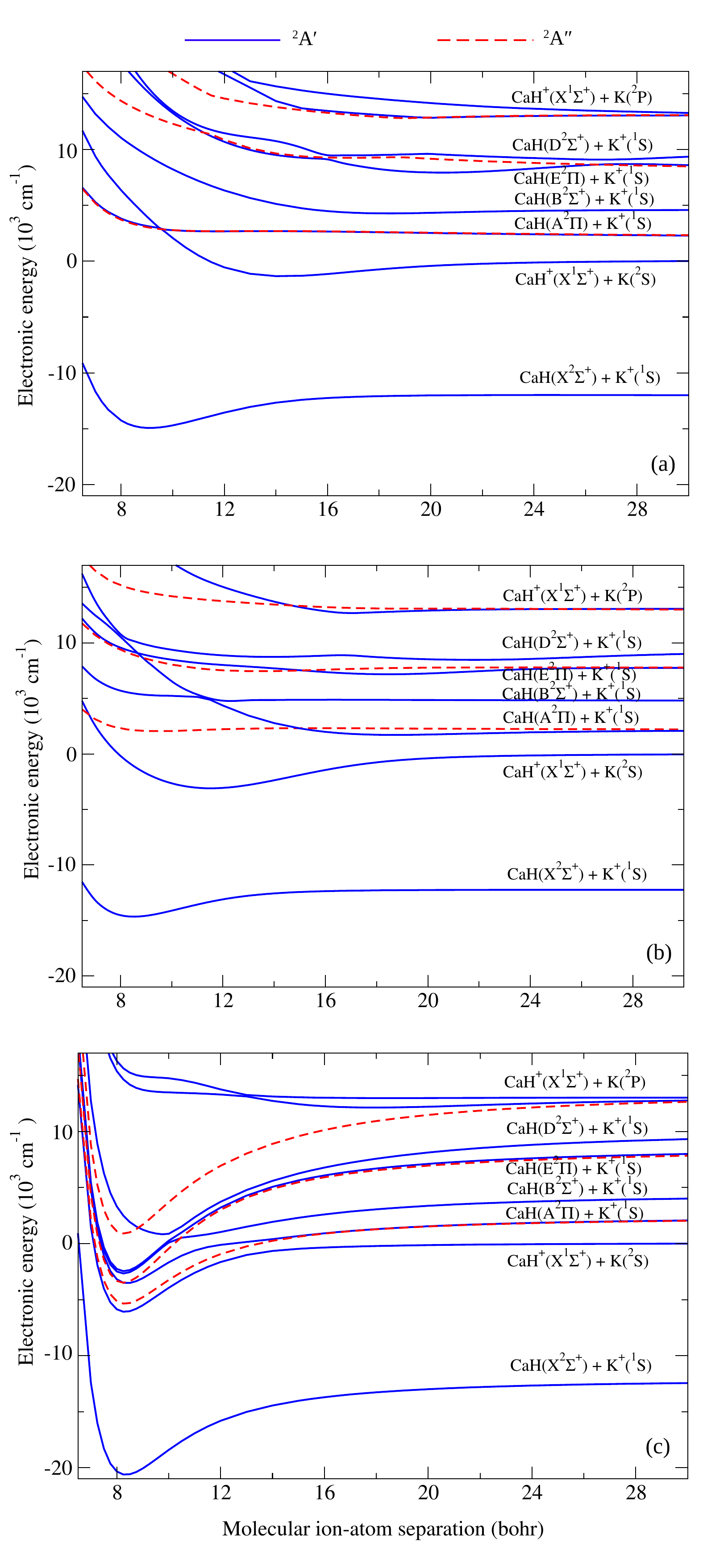}
\caption{\label{fig:S1D}
One-dimensional cuts through the potential energy surfaces for the ground
and excited electronic states of the (CaH-K)$^+$ system for different atom-molecule orientations: (a) $\theta = 0^\circ$ (linear (HCaK)$^+$ configuration), (b) $\theta = 90^\circ$ (T-shape configuration), and (c) $\theta = 170^\circ$ (bent (CaHK)$^+$ configuration).} 
\end{figure}